%% file: ISIT.tex
\documentclass[conference]{IEEEtran}

\usepackage[cmex10]{amsmath}
\usepackage{cases, cite, amsfonts, amssymb, amssymb, amsthm, mathtools, algorithm, algorithmic, psfrag,pstool}
\usepackage{graphicx}
\def\figScale{0.4}

\DeclarePairedDelimiter\abs{\lvert}{\rvert}

\newtheorem{lemma}{Lemma}

\newtheorem{theor}{Theorem}

\begin{document}
\title{On the Achievable Rates of Pairwise Multiway Relay Channels}

\author{\IEEEauthorblockN{Reza Rafie Borujeny}
\IEEEauthorblockA{University of Alberta, AB, Canada\\
Email: reza.rafie@ualberta.ca}
\and
\IEEEauthorblockN{Moslem Noori}
\IEEEauthorblockA{University of British Columbia, BC, Canada\\
Email: moslem@ece.ubc.ca}
\and
\IEEEauthorblockN{Masoud Ardakani}
\IEEEauthorblockA{University of Alberta, AB, Canada\\
Email: ardakani@ualberta.ca}}
\maketitle

\begin{abstract}
\input{Abstract.tex}
\end{abstract}
\IEEEpeerreviewmaketitle

\section{Introduction}\input{Introduction.tex}

\section{Priliminiaries} \label{system} \input{SystemModel.tex}

\section{Problem Definition} \label{problem} \input{Problem.tex}

\section{Problem Solution} \label{solution} \input{Solution.tex}

\section{Simulation Results} \label{simulations} \input{Simulations.tex}

\section{Conclusion} \label{conclusion} \input{Conclusion.tex}

\ifCLASSOPTIONcaptionsoff
  \newpage
\fi
\bibliographystyle{IEEEtran}
\bibliography{MyBib}

\end{document}

%% file: Abstract.tex
In this paper, we study the effect of users' transmission ordering on the common rate and sum rate of pairwise multiway relay channels (MWRCs) with functional-decode-forward strategy. To this end, we first develop a graphical model for the data transmission in a pairwise MWRC. Using this model, we then find the optimal orderings that achieve the maximum common rate and sum rate of the system. The achieved maximum common/sum rate is also found. Moreover, we show that the performance gap between optimal orderings and a random ordering vanishes when SNR increases. Computer simulations are presented for better illustration of the results.

%% file: Introduction.tex
A multiway relay channel (MWRC) \cite{Gunduz2009} is an extension of a two-way relay channel \cite{TWRC1,Wilson,TWRC3,TWRC4,Popovski} in which $N \geq2$ users communicate with each other by means of a relay. There is often no direct link between users and they merely communicate with the relay. Conference calls, file sharing, and multi-player gaming \cite{Moslem} are potential applications of MWRCs. 

Depending on the relay's strategy for forming its downlink message, several relaying schemes have been considered for MWRCs, namely \textit{amplify-and-forward \emph{(AF)}}, \textit{decode-and-forward \emph{(DF)}}, \textit{compress-and-forward \emph{(CF)}} and \textit{functional-decode-forward \emph{(FDF)}} \cite{Gunduz2009, Ong_Binary}. Among these schemes, FDF is the most recent where instead of decoding users' messages separately, the relay directly decodes a function (commonly the sum) of the users' messages. 

FDF is commonly employed along with a pairwise transmission scheme \cite{Ong_Binary} where  similar to two-way relaying, a pair of users transmit their data simultaneously to the relay in each uplink phase. This is then followed by a downlink phase in which the relay broadcasts a function of the received information in the uplink phase to all users. Pairwise transmissions continue until all users are capable of decoding the data of others. Pairwise relaying not only does have a lower decoding complexity than full decoding, but also possesses interesting capacity-achieving properties in different setups \cite{Ong_Binary,Cadambe,Ong,Katti}.

In a pairwise MWRC, the way that users are paired for transmission is referred to as \emph{user's ordering}. As argued in \cite{Moslem}, for an asymmetric MWRC, this ordering directly affects the achievable data rates of the users. To this end, the authors find the optimal ordering to maximize the achievable common rate of the users for an MWRC with asymmetric Gaussian channels under the assumption that each user transmits in at most two uplink phases. For relaying strategy, they considered pairwise FDF and DF relaying and show that the optimal ordering for each strategy is different than the other.

In this work, we go one step further than the work in \cite{Moslem} and address the effect of ordering for a more general pairwise MWRC scenario. More precisely, we consider a pairwise FDF scenario where there is no restriction on the number of uplink transmissions by the users. In this case, we first discuss that there exist $N^{N-2}$ distinct orderings which makes finding the optimal ordering through brute-force search expensive for large $N$. Then, under a reasonable assumption on user's SNR, we analytically find the optimal orderings for the common rate and the sum rate. Using the optimal ordering, we find the maximum achievable common and sum rates. Further, we study the asymptotic behavior of the sum rate for high SNR. This reveals that a randomly chosen ordering performs well for high SNR regimes while the significance of our proposed optimal orderings is more pronounced in low SNRs. 

The rest of the paper is organized as follows. In Section \ref{system}, we describe the system model and introduce a novel graphical interpretation for data transmission in pairwise MWRCs. The sum rate and common rate maximization problems for FDF MWRC are described in Section \ref{problem}. The solution to these problems along with the asymptotic study of the sum rate is presented in Section \ref{solution}. We compare the performance of our proposed orderings with those of randomly chosen orderings via simulations in Section \ref{simulations}. Finally, Section \ref{conclusion} concludes the paper.

%% file: SystemModel.tex
\subsection{System Model}
We consider an MWRC with $N$ users, denoted by $U_1, U_2, \dots, U_N$, where each user $U_i$ wants to share its message $X_i$ with other users. Users cannot directly communicate with each other, thus, relay $\cal{R}$ is used to assist them.  The channel from $U_i$ to $\mathcal{R}$ is a half-duplex reciprocal channel denoted by $C_{i\mathcal{R}}$ with gain $g_{i\mathcal{R}}$. Also, transmitted signals are contaminated by a Gaussian noise  with variance $\sigma^2$.

In a pairwise scheme, the users are divided into $M$  pairs which are not necessarily disjoint. A division of the users to subsets of pairs is called an \textit{ordering} of the users and is denoted by $O = \{\{u_{11},  u_{12}\}, \dots, \{u_{M1}, u_{M2}\}\}$ where $u_{i1}, u_{i2} \in \{U_1, U_2, \dots, U_N\}$. The users exchange their data in one communication round consisting of $M$ uplink and $M$ downlink phases. During each uplink phase, users in one of the pairs transmit their data to the relay. After receiving the users' signal, relay directly decodes the sum of their messages \cite{Nam2010} and broadcasts the sum to all users in a downlink phase. 
This means that if $X_i$ and $X_j$ are vectors with elements chosen from a field $\mathbb{F}$, then the relay directly decodes $X_i \oplus X_i$ where $\oplus$ means element-wise summation of $X_i$ and $X_j$ over $\mathbb{F}$. We consider AWGN channels such that $\oplus$ means element-wise summation over real numbers.  These pairwise transmissions continue until the last pair of the ordering. Having its own data, each user is able to decode the data of others at the end of each round. The transmit power of $U_i$ during an uplink phase is assumed to be $P_i$.  That said, a signal to noise ratio  for user $U_i$,  namely $x_i$, is defined as $x_i \triangleq \frac{P_i\abs{g_{i\mathcal{R}}}^2}{\sigma^2}$. Without loss of generality, we assume that $x_N \geq x_{N-1} \geq \dots \geq x_1 > 0$. 

\begin{figure}[!t]
\psfrag{U1}{\scriptsize$U_1$}
\psfrag{U2}{\scriptsize$U_2$}
\psfrag{U3}{\scriptsize$U_3$}
\psfrag{Relay}{\scriptsize$\mathcal{R}$}
\psfrag{11}{\footnotesize Uplink 1}
\psfrag{21}{\footnotesize Uplink 2}
\psfrag{31}{\footnotesize Uplink 3}
\psfrag{12}{\footnotesize Downlink 1}
\psfrag{22}{\footnotesize Downlink 2}
\psfrag{32}{\footnotesize Downlink 3}
\centering
\includegraphics[scale = 0.1]{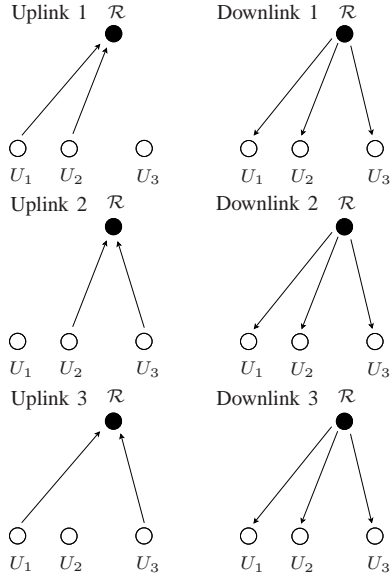}
\caption{A pairwise ordering with $M = N = 3$}
\label{pairwise}
\end{figure}

Fig. \ref{pairwise} illustrates a pairwise MWRC when $N=3$. After a round of communication, each user has the following set of equations:
\begin{equation}
\begin{split}
X_1\oplus X_2 = C_1 
\\X_2\oplus X3 = C_2
\\X_3\oplus X_1 = C_3
\end{split}
\end{equation}
where $C_1$, $C_2$ and $C_3$ are the signals transmitted by the relay.  One can see that the system of equations at each user is solvable using the knowledge of its own data. In a general $N$-user MWRC, if the system of equations at each user is solvable, we say that the corresponding ordering is \textit{feasible}.  This feasibility implies that $M$ should not be less than $N - 1$ because each user needs to find $N-1$ other users' messages. 

In a pairwise MWRC with $M$ pairs, a rate tuple $(R_1, R_2, \dots, R_N)$ is achievable if $U_i$ can reliably (with arbitrarily small probability of error) transmit its data to all other users with rate $R_i$ after each round's $M$ uplink and downlink phases.  The achievable rate tuple depends on the transmit power of the users and the relay as well as the channel gains and the noise power.  Here, we assume that the data rates are limited by the uplink phase, not by the downlink phase. This commonly holds for most wireless systems where users are low-power mobile devices. 

When $U_i$ participates in  a pairwise transmission, say with $U_j$,  during an uplink phase, $R_i$ is limited by the following achievable bound \cite{Nam2010,Moslem}
\begin{align}\label{FDF_rate}
R_i \leq \max\left\{0, \frac{1}{2M}\log_2\left(\frac{x_i}{x_i + x_j}+x_i\right)\right\}.
\end{align}

and to the best of our knowledge, this is the tightest achievable bound for $R_i$ with FDF relaying.
The maximum achievable upper bound on $R_i$ can be found by calculating upper bounds, given by (\ref{FDF_rate}), for $R_i$ over all pairs that $U_i$ is part of and then taking the minimum of these bounds.
In this paper, instead of focusing on the individuals' rates, we study the system common rate and sum rate. For an achievable rate tuple $ (R_1, \dots, R_N)$, the user's common rate, $C_R$,  and the sum rate, $S_R$, are defined as $C_R \triangleq \min_i {R_i}$ and $S_R \triangleq \sum_{i=1}^{N}R_i$. As seen from (\ref{FDF_rate}), the upper bounds on $R_i$'s, and consequently the systems common rate and sum rate depend on the ordering of the users. Our goal in this work is to find the orderings that attains the maximum possible common rate and sum rate in the system. This is discussed in more detail later.

\subsection{Graphical Representation}
Here, we introduce the concept of \textit{client graph} that provides a convenient representation of the users' transmission ordering. This model is later used to find the optimal ordering to maximize $C_R$ and $S_R$.

A client graph $G_O = (V,E)$ for a given pairwise ordering $O$  consists of a set of vertices $V = \{v_1, v_2, \dots, v_N\}$ and a set of edges $E$.  Vertex $v_i$ is associated with $U_i$ and  $v_iv_j \in E$ iff $\{U_i, U_j\} \in O$. If $v_iv_j\in E$, we say $v_j$ is adjacent to $v_i$. The set of adjacent vertices of $v_i$, denoted by $A_i^G$, is called the set of neighbors of $v_i$. Also the degree of node $v_i$ is $deg(v_i) = |A_i^G|$. The adjacency matrix of $G_O(V,E)$, denoted by $\mathcal{A} = (a_{ij})$, is an $N\times N$ matrix in which $a_{ij} = 1$ iff $v_iv_j\in E$, and $a_{ij} $ is $0$ otherwise. Note that there is a one-to-one mapping between all possible client graphs and all possible orderings.

The overall energy consumed in a communication round is directly proportional to the number of pairs. As a result,  we are interested in identifying feasible orderings with minimum number of pairs which, as we mentioned, is $M=N-1$. To this end, we state the following theorem.
\begin{theor}\label{th1}
An ordering with $M=N-1$ pairs is feasible iff the corresponding client graph is a tree.
\end{theor}

\begin{IEEEproof}
Fist, we show that if there is a cycle in the client graph $G_O$, the feasibility of the system will not change if we remove one of the edges from that cycle.
Assume that $\mathcal{C} = \{v_{i_1}v_{i_2}, v_{i_2}v_{i_3}, \dots , v_{i_n}v_{i_1}\}$ is a cycle in $G_O$. The equations corresponding to the edges in this cycle are:
\begin{IEEEeqnarray}{c}\label{syseq}
X_{i_1} \oplus X_{i_2} = C_1\IEEEnonumber\\
X_{i_2} \oplus X_{i_3} = C_2\\\IEEEnonumber
\vdots \\\IEEEnonumber
X_{i_n} \oplus X_{i_1} = C_n\IEEEnonumber .
\end{IEEEeqnarray}
The $j$th equation in the system of equations (\ref{syseq}) is not independent of the others. In other words, if we sum over all of the equations but the $j$th one, we wind up with:
\begin{equation}
X_{i_j} \oplus X_{i_{j+1}} = \bigoplus_{i\neq j}{C_i} .
\end{equation}
This shows that removing $v_{i_j}v_{i_{j+1}}$ from the cycle $\mathcal{C}$, has no effect on the feasibility of the system of equations (\ref{syseq}).

Then, we just need to prove the theorem for client graphs with no cycle. In order for system of equations to be feasible, each user needs to have at least $N-1$ equations, except its own data. It means that $G_O$ has at least $N-1$ edges. Since $G_O$ has no cycle, it should be a tree. 
\end{IEEEproof}
In the rest of this paper, we assume $M=N-1$ and use the terms client tree and client graph, interchangeably.

%% file: Problem.tex
In this section, we define rate maximization problems. Here, we denote the maximum achievable common rate and sum rate for a client graph $G_O$ by $C_R(G_O)$ and $S_R(G_O)$, respectively.

By \emph{common rate maximization} problem, we mean finding the feasible ordering that maximizes $C_R(G_O)$. More formally, if we denote the set of all feasible orderings by $\mathcal{O}$, then the common rate maximization problem translates into
\begin{equation}\label{GenCR}
O_{CR} = \operatorname*{arg\,max}_{O \in \mathcal{O}}{C_R(G_O)}
\end{equation}
Similarly, a \emph{sum rate maximization} is defined as follows
\begin{equation}\label{GenSR}
O_{SR}=\operatorname*{arg\,max}_{O \in \mathcal{O}}{S_R(G_O)}
\end{equation}

One way to solve the aforementioned problems is to search over all possible client trees and find the one that maximizes  the common rate and sum rate. This, according to Cayley's formula \cite{cayley}, necessitates searching over all $N^{N-2}$ feasible client trees which is impractical even if the number of users is not very large. This motivates us to find efficient solutions for identifying the optimal client trees.

In order to find an ordering with maximum sum rate, we consider the case where the user's SNR is not too low which is the case for most practical settings. To this end, the upper bound on the rate of $U_i$ when it transmits with $U_j$ is given by
\begin{equation}\label{FDF_weak}
R_i \leq \frac{1}{2(N-1)}\log_2\left(\frac{x_i}{x_i + x_j}+x_i\right).
\end{equation}
One can easily verify that if $x_1 + \frac{x_1}{x_1 + x_N} \geq 1$, (\ref{FDF_weak}) and (\ref{FDF_rate}) are equivalent. For instance, if all SNRs of the users are more than 1, the bound in (\ref{FDF_weak}) is equivalent to (\ref{FDF_rate}).
For common rate maximization, we also assume that the user's SNR is not too low and consider (\ref{FDF_weak}). We are not interested in cases that common rate is equal to zero.

%% file: Solution.tex
In this section, we provide solutions to common rate and sum rate maximization problems for FDF relaying. We also show that in high SNR regimes, the performance of a randomly chosen ordering asymptotically approaches the rate performance of the optimal ordering.

\subsection{Common Rate Maximization}
\begin{theor}\label{FDFCR}
$C_R(G_O)$ is maximized when the ordering is
\begin{equation} \nonumber
O_{CR} \!=\! \{\{U_1 , \!U_2\},\{U_2,\!U_3\}, \{U_3,\!U_4\},\dots ,\{U_{N-1},\!U_N\}\}.
\end{equation}
Also, the maximum achievable common rate is
\begin{equation} \nonumber
C_R(G_O) \! = \! \! \min_{i \in \{1,\ldots,N\}} \! \! \left\{ \! \! \frac{1}{2(N-1)}\log_2\left( \! x_i \! +\!  \frac{x_i}{x_i \! +\! x_{i+1}} \! \right)  \! \! \right\}.
\end{equation}
\end{theor}

\begin{figure}[!t]
\psfrag{U_1}{$U_1$}
\psfrag{U_2}{$U_2$}
\psfrag{U_3}{$U_3$}
\psfrag{U_N}{$U_{N}$}
\psfrag{...}{$\dots$}
\centering
\includegraphics[scale = 0.3]{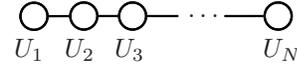}
\caption{Client tree that maximizes $C_R(G_O)$ for a pairwise MWRC with FDF relaying}
\label{CRFDFgraph}
\end{figure}

\begin{IEEEproof}
Here, by an optimal tree, we mean a client tree that achieves the maximum $C_R$ with respect to (\ref{FDF_rate}). There are two statements regarding (\ref{FDF_rate}) which we use to prove the theorem:
\begin{enumerate}
\item The function $f(x) = x\left(1+\frac{1}{x+ \alpha}\right)$ is an increasing function of $x$.
\item The function $g(x) = \left(1+\frac{1}{\alpha + x}\right)$ is a decreasing function of $x$.
\end{enumerate}
Given a client tree, $G_O(V,E)$, with an FDF MWRC, we have
\begin{equation}\label{CRFDFdef}
C_R(G_O) = \min_{i,j} \left\{ \frac{1}{2(N-1)}\log_2\left(x_i+ \frac{x_i}{x_i + x_j}\right)  \right\}.
\end{equation}
where $\ x_i \leq x_j$ and $ v_iv_j \in E$. Using (\ref{CRFDFdef}), we prove the following lemma. 

\begin{lemma}\label{FDFCRlemma}
There exists an optimal tree ,$G_O(V,E)$, in which $A_1^{G_O} = \{v_2\}$.
\end{lemma}
\begin{IEEEproof}
We adapt $G_{O'}(V,E')$ from $G_{O}$ such that we disconnect all of the neighbors of $v_1$ from $v_1$ and connect them to $v_2$. We also make $v_1$ and $v_2$ neighbors. More precisely,
\begin{equation}
E' = (E - \{v_1v_i|v_i \in A_1^{G_O}\}) \cup \{v_2v_i|v_i \in A_1^{G_O}; i \neq 2\} \cup \{v_1v_2\}
\end{equation}
Because of monotonicity of $f(x)$ and $g(x)$, to verify that $C_R(G_O) \leq C_R(G_{O'})$ we just need to show
\begin{equation}\label{mideqn}
x_1\left(1+\frac{1}{x_1 + x_{min}}\right) \leq x_2\left(1+\frac{1}{x_2 + x1}\right)
\end{equation}
where, $x_{min} = \min \{x_i| v_i \in A_1^{G_O}\}$. After some manipulation, we find that (\ref{mideqn}) is equivalent to
\begin{equation}
0 \leq (x_2-x_1)(x_1+x_{min})(x_2 + x_1) + x_2x_{min} - x_1^2
\end{equation}
which, according to the fact that $x_1 \leq x_{min}$, is true.
\end{IEEEproof}
We prove the theorem by induction. If $N = 2$ the theorem obviously holds. Now, assume that the statement of the theorem holds for every FDF MWRC with $N=k$. We show that it also holds for any FDF MWRC with $N=k+1$. For $N=k+1$, according to Lemma (\ref{FDFCRlemma}), there exists an optimal tree $G_O(V,E)$ in which $A_1^{G_O} = \{v_2\}$. From equation (\ref{CRFDFdef}), we also have:
\begin{align}\label{CRGO} \nonumber
C_R(G_O\!) \!  = \! & \min_{  i,j}  \! \! \left\{ \frac{1}{2(N \!\!-\!\!  1\!)} \!  \log_2 \! \! \left( \! x_i \! +\! \! \frac{x_i}{x_i \! +\!  x_j} \!\!\right)\!\!|  \! 1 \! \! < \! i \! \leq \! \!  j ;  v_iv_j \!\!  \in \! \! E \! \right \}\\ 
&\cup \left\{\frac{1}{2(N-1)}\log_2\left(x_1+ \frac{x_1}{x_1 + x_2}\right) \right\}
\end{align}
If the second term in (\ref{CRGO}) is the limiting term in all of the possible client trees with $A_1^{G_O} = \{v_2\}$, the proposed ordering is optimal. Otherwise, maximizing $C_R(G_O)$ is equivalent to maximizing $\min \left\{x_i\left(1+ \frac{1}{x_i + x_j}\right)|1 < i \leq j ; v_iv_j \in E \right\}$. It is equivalent to maximizing the $C_R$ for $G_{O'}(V',E')$, in which $V' = V - \{v_1\}$ and $E' = E - \{v_1v_m| v_m \in A_1^{G_O}\}$. According to the induction hypothesis, it happens when $O' = \{\{v_2v_3\}, \{v_3v_4\}, \dots , \{v_{N-1}v_N\}\}$ and as a reslut
\begin{equation}
O = \{\{v_1v_2\}, \{v_2v_3\}, \dots , \{v_{N-1}v_N\}\}
\end{equation}
\end{IEEEproof}

\begin{figure}[!t]
\psfrag{U_1}{$U_1$}
\psfrag{U_2}{$U_2$}
\psfrag{U_3}{$U_3$}
\psfrag{U_N}{$U_N$}
\psfrag{...}{$\dots$}
\centering
\includegraphics[scale = 0.3]{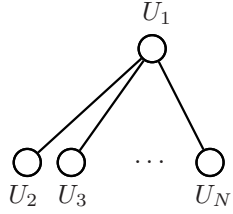}
\caption{Client tree that maximizes $S_R(G_O)$ for a pairwise MWRC with FDF relaying  subject to the weakened upper bound given by (\ref{FDF_weak})}
\label{SRgraph}
\end{figure}

Fig. \ref{CRFDFgraph} illustrates the optimal ordering for an FDF MWRC that achieves the maximum $C_R$.

\subsection{Sum Rate Maximization}
\begin{theor}\label{FDFSR}
 $O = \{\{U_2,U_1\},\{U_{3},U_1\},\dots ,\{U_{N},U_1\}\}$ is the optimal ordering maximizing the sum rate subject to (\ref{FDF_weak}). Moreover, the maximum sum rate for this ordering is
\begin{align}\label{maxSRFDF} \nonumber
S_R(G_O) \!= \!\frac{1}{2(N\!-\!1)} \log_2 & \Bigg( \max \left\{1, \left(x_1 + \frac{x_1}{x_1 + x_N}\right) \right\}\\  
& \times \! \!  \prod_{i=2}^{N} \! {\max \left\{1, \frac{x_i}{x_i \! +\!  x_1} \! + \! x_i \! \right\}}  \! \Bigg).
\end{align}
\end{theor}
\input{FDFSR.tex}

Fig. \ref{SRgraph} illustrates the optimal ordering for an FDF MWRC that achieves the maximum $S_R$.

\begin{figure}[!t]
\centering
\input{Figure1_corrected.tex}
\includegraphics[scale = \figScale]{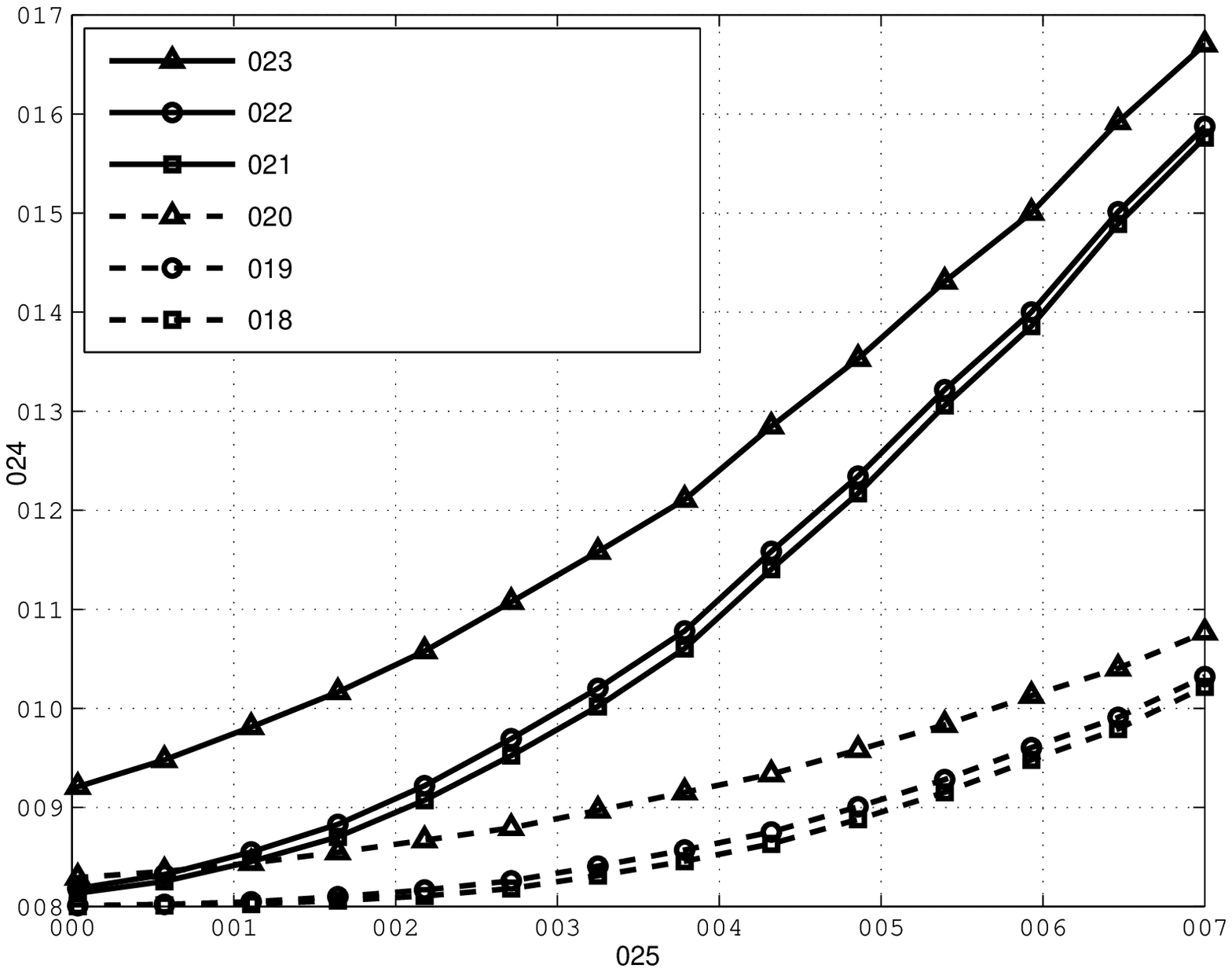}
\caption{Comparison between the common rate of the optimal ordering and random ordering in MWRC with FDF relaying for $N = 4$ and $8$}
\label{simFDFCR}
\end{figure}

\subsection{Asymptotic Behavior}
Using Theorem \ref{FDFCR}, it is straightforward to show that
\begin{equation}
C_R(G_O) - C_R(G_{O'}) \leq \frac{1}{2(N-1)}\log_2{\left({\frac{1+2x_N}{2x_1}}\right)}
\end{equation}
where $O$ and $O'$ refer to the optimal ordering and a random ordering, respectively. Now, if $x_N  \sim x_1 $\footnote{$f(x)$ is on the order of $g(x)$, $f(x) ~\sim g(x)$, if the asymptotic limit of their ratio approaches 1.}, one can conclude that
\begin{equation}\label{asympCR}
\lim_{x_1 \to \infty}{\left(C_R(G_O) - C_R(G_{O'})\right)} = 0.
\end{equation}
Similarly, for high SNR regimes, we have 
\begin{equation}
S_R(G_O) - S_R(G_{O'}) \leq \frac{1}{2}\log_2{\left({\frac{x_N(1+2x_1)}{x_1(1+2x_N)}}\right)}
\end{equation}
and consequently
\begin{equation}\label{asympSR}
\lim_{x_1\to \infty}{\left(S_R(G_O) - S_R(G_{O'})\right)} = 0.
\end{equation}
In summary, equations (\ref{asympCR}) and (\ref{asympSR}) show that for FDF relaying, the performance of a randomly chosen ordering approaches the one for optimal ordering in high SNR regimes.

%% file: FDFSR.tex
To prove the theorem, we first show that there is an optimal tree with $deg(v_N)=1$ (Lemma (\ref{FDFSR1})). Then we prove that in the optimal tree each node needs to have only one neighbor among nodes with a lower SNR (Lemma (\ref{FDFSR2})). We then show that there exist an optimal tree with $deg(v_N) = deg(v_{N-1}) = 1$ (Lemma (\ref{FDFSR4})). In the next step, we prove that in an optimal tree for two nodes of degree one, say $v_i$ and $v_j$, if $v_i$ has a higher SNR than $v_j$ then the neighbor of $v_i$ has a higher SNR than the neighbor of $v_j$ (Lemma (\ref{FDFSR5})). Then we prove the theorem by induction (Lemma (\ref{FDFSR6}))
\begin{IEEEproof}
We use the following convention for the rest of this proof:
\begin{equation}
d_i \triangleq 2^{2(N-1)R_i}.
\end{equation}
As a result, the bound given by (\ref{FDF_weak}) is equivalent to
\begin{equation}
d_i \leq x_i\left(1 + \frac{1}{x_i+x_j}\right).
\end{equation}
We also define $D_s(G_O) = \max\ \prod_{i=1}^{N}{d_i} = 2^{2(N-1) S_R(G_O)}$. Assume that $G(V,E)$ is a tree such that $\{v_i, v_j, v_k\} \subseteq V$ and$\{v_iv_j, v_iv_k\} \subseteq E $. We define a $V$-transform on $G$ in such a way that $V(G,v_i,v_j,v_k) = G'(V, E')$ and $E' = (E - \{v_iv_k\}) \cup \{v_jv_k\}$. Fig. \ref{Vtran} shows the operation of a $V$-transform.
\begin{figure}[!t]
\psfrag{U_1}{$v_j$}
\psfrag{U_2}{$v_i$}
\psfrag{U_3}{$v_k$}
\psfrag{...}{$\dots$}
\psfrag{arrow}{$\rightarrow$}
\centering
\includegraphics[scale = 0.3]{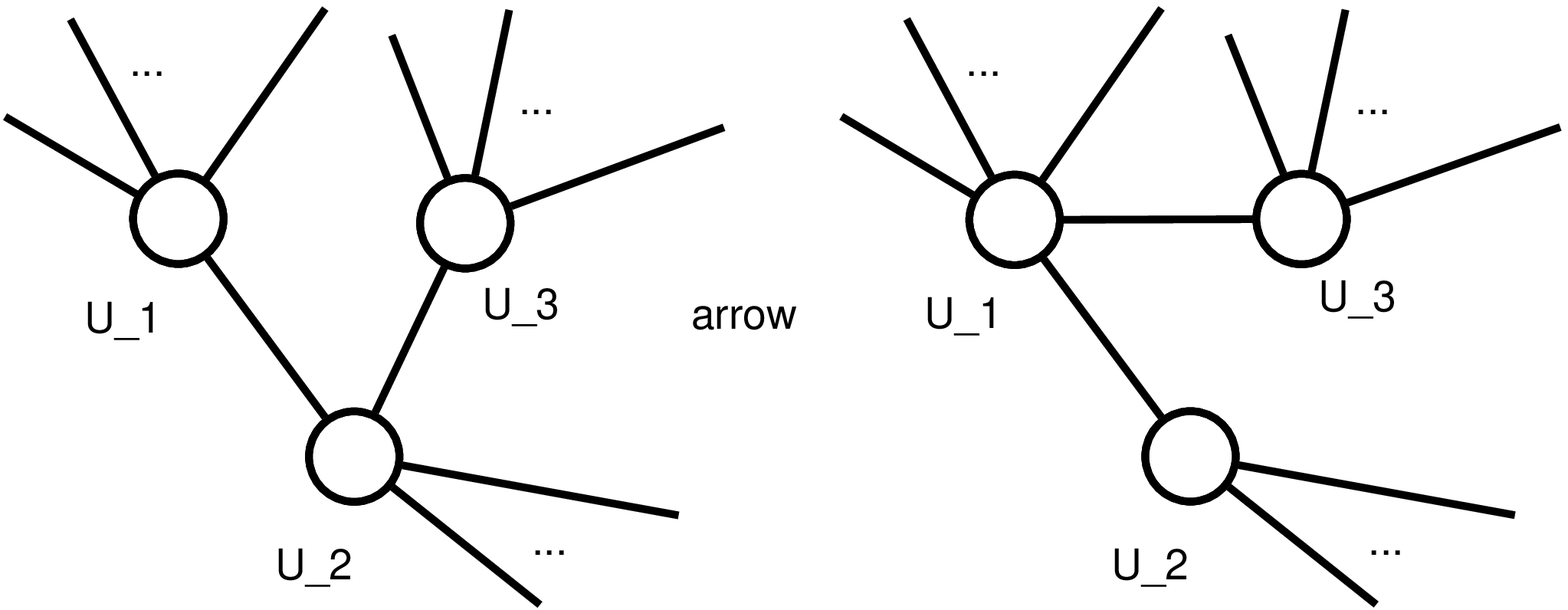}
\caption{Operation of $V$-transform, $V(G, v_i,v_j,v_K)$}
\label{Vtran}
\end{figure}
\begin{lemma}\label{FDFSR1}
There exists an optimal tree in which $deg(v_N) = 1$.
\end{lemma}
\begin{IEEEproof}
Assume $G_O$ is an optimal tree in which $deg(v_N)>1$ and $v_i$ and $v_j$ are two neighbors of $v_N$ and $x_j$ is the minimum SNR value of the neighbors of $V_n$. Consequently, we have $x_i \geq x_j$. It is straightforward to show that by performing a $V$-transform on $G_O$ and transform it to $G_{O'} = V(G_O,v_N,v_i,v_j)$,  we have $\frac{D_s(G_{O'})}{D_s(G_O)} \geq 1$. It means that the sum rate of $G_{O'}$ is not less than sum rate of $G_O$.
Note that, after applying this $V$-transform, we have reduced degree of $v_N$ by one. After applying $deg(v_N) - 2$ more $V$-transforms, we end up with an optimal tree with $deg(v_N) = 1$. Fig. \ref{Vdeg4} illustrates an hypothetical optimal tree with $deg(v_N) = 4$. It shows how we apply $3$ $V$-transforms to get an optimal tree with $deg(v_N) = 1$.
\begin{figure}[!h]
\psfrag{U_2}{$v_N$}
\psfrag{U_3}{$v_j$}
\psfrag{...}{$\dots$}
\psfrag{arrow}{$\rightarrow$}
\centering
\includegraphics[scale = 0.3]{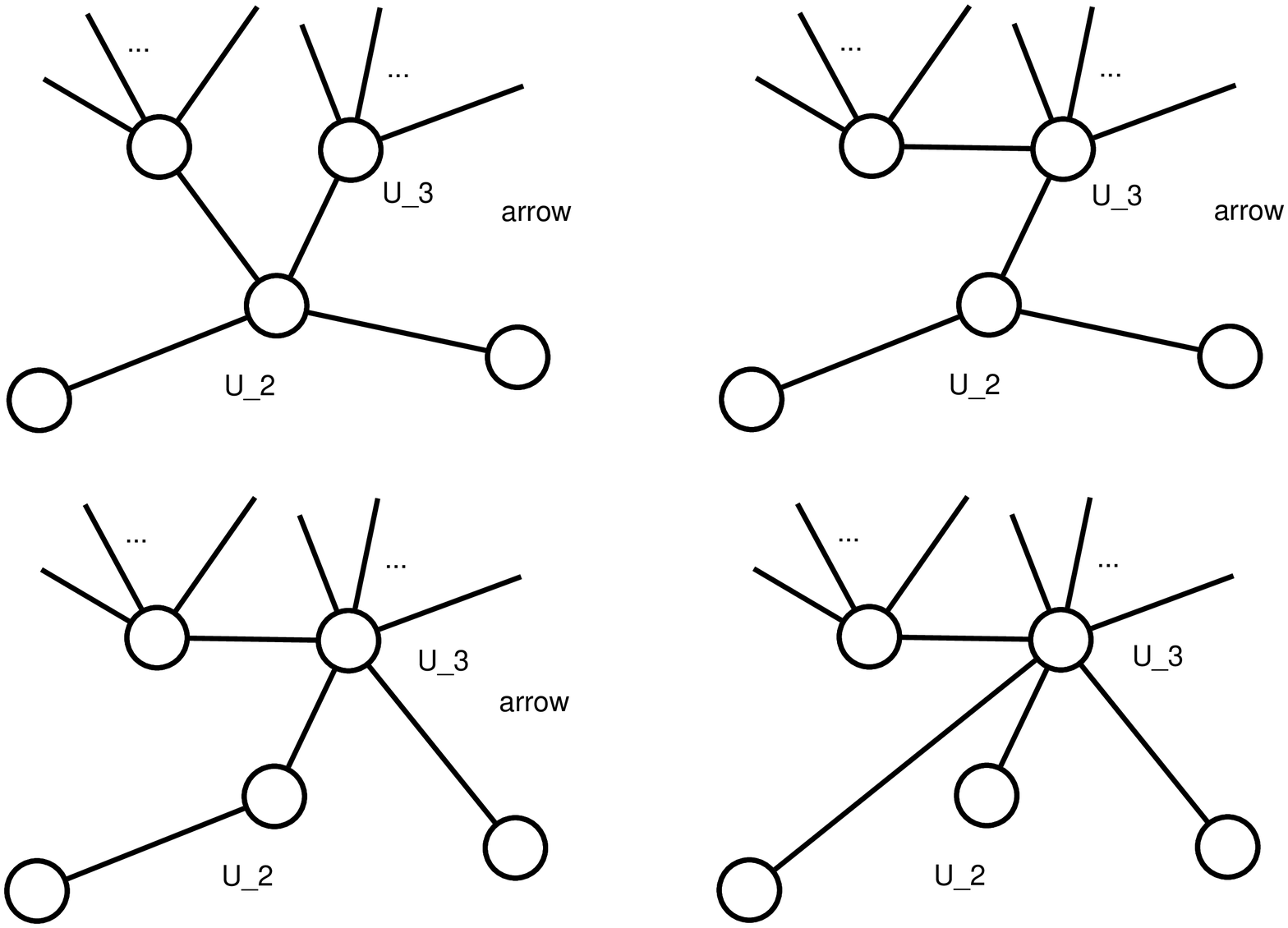}
\caption{Applying 3 $V$-transform on an optimal tree with $deg(v_N) = 4$}
\label{Vdeg4}
\end{figure}
\end{IEEEproof}

\begin{lemma}\label{FDFSR2}
There exists an optimal tree, $G_O(V,E)$, such that for any $0<i<N-1$, $deg(v_{N-i}) \leq i+1$.Furthermore, the number of neighbors of $v_{N-i}$ with a lower SNR than $V_{N-i}$ is at most one and consequently, the number of neighbors of $v_{N-i}$ which have higher SNR than $x_{N-i}$ is at least $deg(v_{N-i})-1$.
\end{lemma}
\begin{IEEEproof}
If the number of those neighbors of $v_{N-i}$ that have a lower SNR value than $x_{N-i}$ is $a$, after applying $a-1$ $V$-transforms, we end up with an optimal tree in which $deg(v_{N-i}) \leq i+1$. These $a-1$ $V$-transforms have the form $V(G,V_{N-i},v_i,v_k)$ and $v_k$ has the highest SNR value among all of the neighbors of $v_{N-i}$.

Now, assume that $deg(v_{N-i}) \leq i+1$ and $v_{N-i}$ has at most one neighbor $v_j$ such that $j < N-i$. Then, we have that the number of neighbors of $v_{N-i}$ that have a higher SNR than $x_{N-i}$ is $\geq |A_{N-i}^{G_O}|-1 = deg(v_{N-i})-1$.
\end{IEEEproof}

\begin{lemma}\label{FDFSR4}
There exists an optimal tree, $G_O(V,E)$, in which $deg(v_{N}) = deg(v_{N-1}) = 1$. Moreover, if $v_j$ is the only neighbor of $v_{N-1}$ and $v_i$ is the only neighbor of $v_{N}$, then $x_i \geq x_j$
\end{lemma}
\begin{IEEEproof}
If $deg(v_{N-1}) = 2$, according to Lemma (\ref{FDFSR2}) and (\ref{FDFSR1}), there exists an optimal tree $G_O(V,E)$ in which $deg(v_{N})=1$ and $v_Nv_{N-1} \in E$. Let the other neighbor of $v_{N-1}$ be $v_j$. Then, $G_{O'} = V(G_O,v_N,v_i,v_j)$ is an optimal tree in which $deg(v_{N-1}) = 1$. So, there always exists an optimal tree $\ G_O$, with $deg(v_{N}) = deg(v_{N-1}) = 1$. Assume that the only neighbor of $v_{N-1}$ is $v_{j}$. If $v_{j} = v_{N}$, the graph will be disconnected. Otherwise, if the only neighbor of $v_N$ is $v_i$, we want to prove that $x_i \geq x_j$. We also assume $x_N \neq x_{N-1}$; otherwise, one can rename the nodes in such a way that theorem holds. Assume that $G_{O''}(V,E'')$ is a client tree in which:
\begin{equation}
E'' = (E - \{v_Nv_i, v_{N-1}v_j\}) \cup \{v_Nv_j, v_{N-1}v_i\}.
\end{equation}
It is easy to show that $D_s(G_{O''}) \leq D_s(G_{O})$ iff $x_i \geq x_j$.
\end{IEEEproof}
Next lemma, is a generalization of Lemma (\ref{FDFSR4}) and we prove it in a similar way. It roughly says that in an optimal tree a node with a higher SNR has a neighbor with a higher SNR.
\begin{lemma}\label{FDFSR5}
Assume that $G_O(V,E)$ is an optimal tree in which $deg(v_{N}) = deg(v_{N-1}) = \dots = deg(v_{N-i}) = 1$ and $i < N - 1$. Also, assume that $q < p \leq i$ and $\{v_jv_{N-p}, v_kv_{N-q}\} \in E$. Then $x_j \leq x_k$.
\end{lemma}
\begin{IEEEproof}
It is obvious that $j> N - i$ and $k > N - i$, otherwise the graph is disconnected. Now, if $x_k < x_j$, according to Lemma (\ref{FDFSR4}), the graph $G_{O'}(V, E')$ with $E' = (E - \{v_jv_{N-p},v_kv_{N-q}\}) \cup \{v_jv_{N-q}, v_kv_{N-p}\}$ has a greater sum rate which contradicts the fact that $G_O$ is optimal.
\end{IEEEproof}

\begin{lemma}\label{FDFSR6}
Assume $G_O(V,E)$ is an optimal tree and $i$ is the largest integer that 
\begin{equation}
deg(v_N) = deg(v_{N-1}) = \dots = deg(v_{N-i}) = 1.
\end{equation}
If $ i < N-1$, then there exists an optimal tree $G_{O'}(V,E')$ in which 
\begin{equation}
deg(v_N) = deg(v_{N-1}) = \dots = deg(v_{N-i+1}) = 1.
\end{equation}
\end{lemma}
\begin{IEEEproof}
Assume that $A_{N-i+1}^{G_O} \cap \{v_N, v_{N-1}, \dots, v_{N-i}\} = \{v_{m_1}, v_{m_2}, \dots, v_{m_n}\}$ where $m_1 > m_2 > \dots > m_n$. Define 
\begin{equation}
B = A_{N-i+1}^{G_O} - \{v_N, v_{N-1}, \dots, v_{N-i}\}.
\end{equation}
According to Lemma (\ref{FDFSR2}), we assume that $|B| \leq 1$. If $|B| = 0$, $G_O$ is disconnected. Assume $B = \{v_j\}$. Consider $G_{O'}(V,E')$ such that 
\begin{align}\nonumber
E' = (E -  \{v_{m_1}v_{N-i+1}, v_{m_2}v_{N-i+1}, \dots, v_{m_n}v_{N-i+1}\})\\ \cup \{v_{m_1}v_j, v_{m_2}v_j, \dots, v_{m_n}v_j\}.
\end{align}
Then, one can conclude that $\frac{D_s(G_O)}{D_s(G_{O'})} \geq 1$.
\end{IEEEproof}

According to Lemma (\ref{FDFSR6}), there exists an optimal tree with respect to (\ref{FDF_weak}) in which
\begin{equation}
deg(v_N) = deg(v_{N-1}) = \dots = deg(v_{2}) = 1.
\end{equation}
As a result, $O$ is an optimal solution with respect to (\ref{FDF_weak}). The muximum achievable sum rate, $S_R(G_O)$, could be found directly from (\ref{maxSRFDF}).
\end{IEEEproof}

%% file: Figure1_corrected.tex
%
%
\providecommand\matlabtextA{\color[rgb]{0.000,0.000,0.000}\fontsize{6}{15}\selectfont\strut}%
\providecommand\matlabtextE{\color[rgb]{0.000,0.000,0.000}\fontsize{6}{35}\selectfont\strut}%
\providecommand\matlabtextB{\color[rgb]{0.000,0.000,0.000}\fontsize{6}{10}\selectfont\strut}%
\providecommand\matlabtextC{\color[rgb]{0.000,0.000,0.000}\fontsize{5}{10}\selectfont\strut}%
\providecommand\matlabtextD{\color[rgb]{0.000,0.000,0.000}\fontsize{5}{10}\selectfont\strut}%
\psfrag{018}[cl][cl]{\matlabtextB Random ordering, $N = 8$}%
\psfrag{019}[cl][cl]{\matlabtextB Optimal ordering, $N = 8$}%
\psfrag{020}[cl][cl]{\matlabtextB Upper bound, $N = 8$}%
\psfrag{021}[cl][cl]{\matlabtextB Random ordering, $N = 4$}%
\psfrag{022}[cl][cl]{\matlabtextB Optimal ordering, $N = 4$}%
\psfrag{023}[cl][cl]{\matlabtextB Upper bound, $N = 4$}%
\psfrag{024}[bc][bc]{\matlabtextE $C_R$ (bits of information per MWRC phase)}%
\psfrag{025}[tc][tc]{\matlabtextA ${1} / {\sigma^2}$ (dB)}%
%
%
%
\def\matlabfragNegXTick{\mathord{\makebox[0pt][r]{$-$}}}
\psfrag{000}[ct][ct]{\matlabtextC $1$}%
\psfrag{001}[ct][ct]{\matlabtextC $3$}%
\psfrag{002}[ct][ct]{\matlabtextC $5$}%
\psfrag{003}[ct][ct]{\matlabtextC $7$}%
\psfrag{004}[ct][ct]{\matlabtextC $9$}%
\psfrag{005}[ct][ct]{\matlabtextC $11$}%
\psfrag{006}[ct][ct]{\matlabtextC $13$}%
\psfrag{007}[ct][ct]{\matlabtextC $15$}%
%
%
%
\psfrag{008}[rc][rc]{\matlabtextD $0$}%
\psfrag{009}[rc][rc]{\matlabtextD $0.05$}%
\psfrag{010}[rc][rc]{\matlabtextD $0.1$}%
\psfrag{011}[rc][rc]{\matlabtextD $0.15$}%
\psfrag{012}[rc][rc]{\matlabtextD $0.2$}%
\psfrag{013}[rc][rc]{\matlabtextD $0.25$}%
\psfrag{014}[rc][rc]{\matlabtextD $0.3$}%
\psfrag{015}[rc][rc]{\matlabtextD $0.35$}%
\psfrag{016}[rc][rc]{\matlabtextD $0.4$}%
\psfrag{017}[rc][rc]{\matlabtextD $0.45$}%
%

%% file: Simulations.tex
\begin{figure}[t]
\centering
\input{Figure2_corrected.tex}
\includegraphics[scale = \figScale]{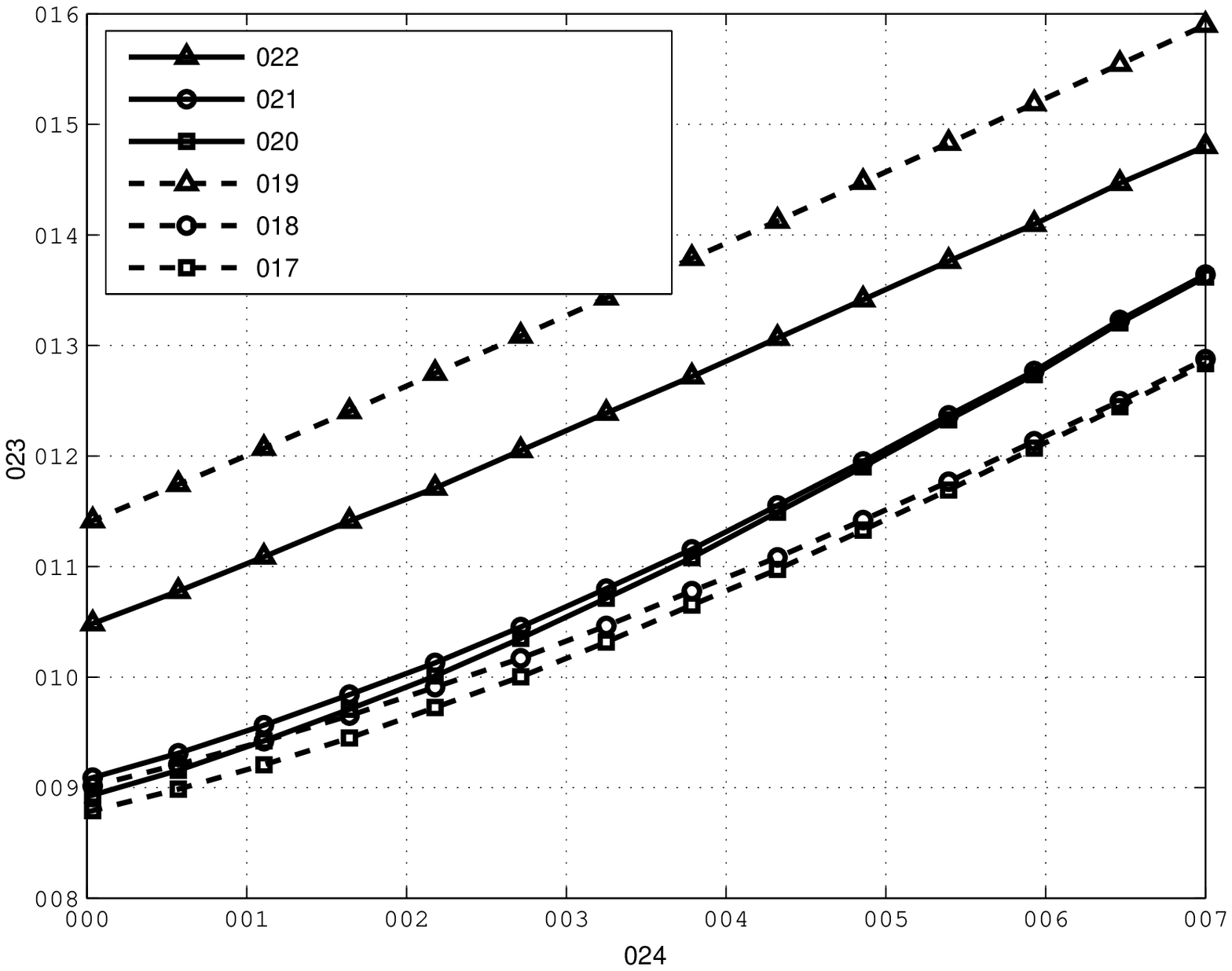}
\caption{Comparison between the sum rate of the optimal ordering and random ordering in MWRC with FDF relaying for $N = 4$ and $8$}
\label{simFDFSR}
\end{figure}

In this section, we investigate the performance of the optimal ordering in comparison with random
orderings. We use Monte Carlo simulation to compare the optimal
ordering and a randomly selected ordering. For each simulation round,
random ordering is selected uniformly at random from all of the feasible client trees. We again assume that the data rates are limited by the uplink phase. Similar to \cite{Moslem}, it is assumed
that the channels between the users and the relay are Rayleigh fading with parameter 1.  The number
of users is set to $N = 4$ and $8$. In order to illustrate the difference between optimal ordering and
random orderings, we define the \textit{common rate gap} \cite{Moslem} of random ordering and optimal
ordering as $G_C = \frac{C_R(G_O) - C_R(G_{O'})}{C_R(G_O)}$ where, by abuse of notation, we denote the
average of common rate over all of the simulation rounds by $C_R (\cdot)$.  The subscripts $O$ and $O'$
denote optimal ordering and randomly chosen orderings, respectively. Similarly, we define the
\textit{sum rate gap} as $G_S = \frac{S_R(G_O) - S_R(G_{O'})}{S_R(G_O)}\ $. 

\begin{figure}[!h]
\centering
\input{Figure3_corrected.tex}
\includegraphics[scale = \figScale]{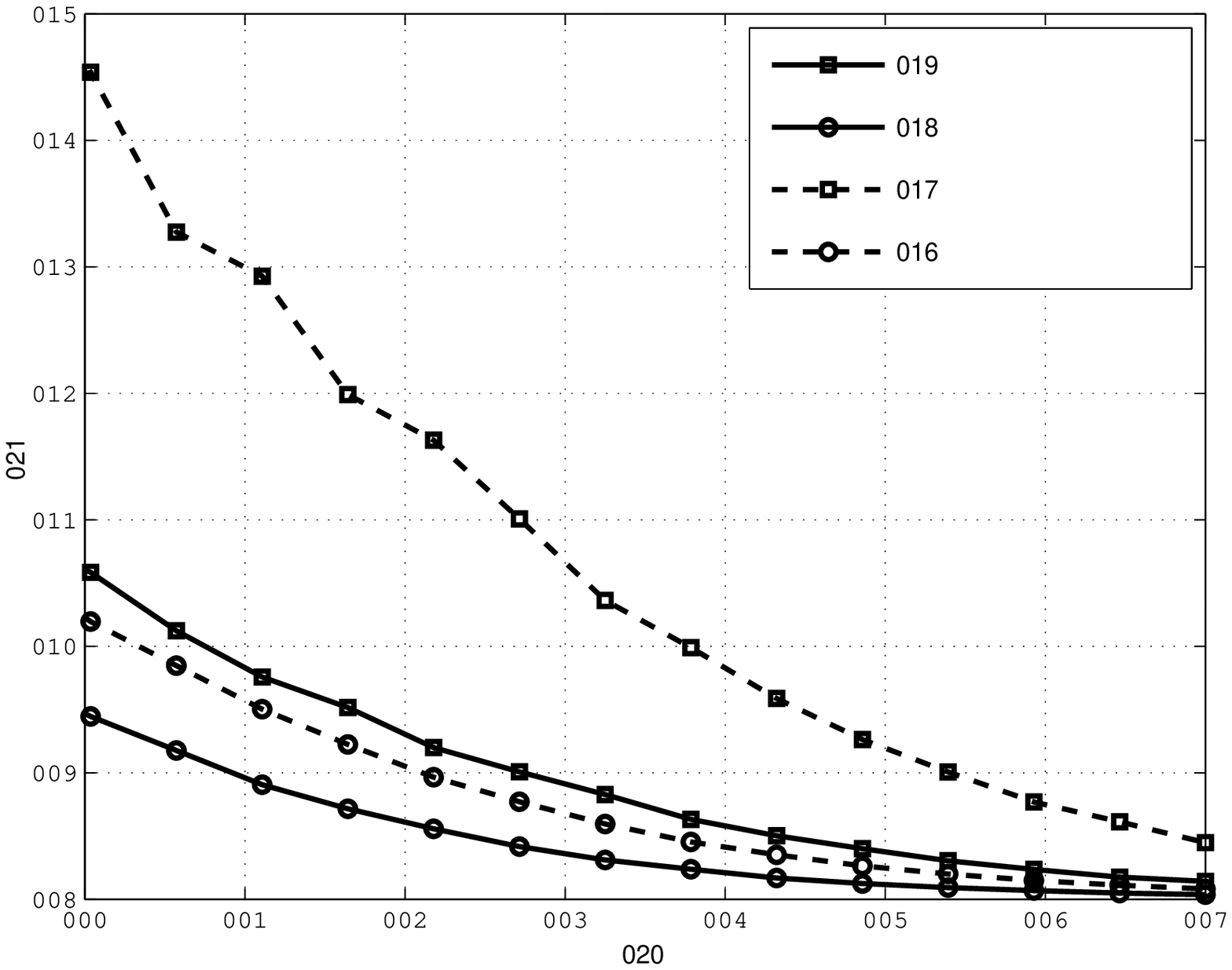}
\caption{Common rate and sum rate gap between optimal ordering and random ordering for $N = 4$ and $8$}
\label{GCS}
\end{figure}

\begin{figure}[!h]
\centering
\input{Figure4_corrected.tex}
\includegraphics[scale = \figScale]{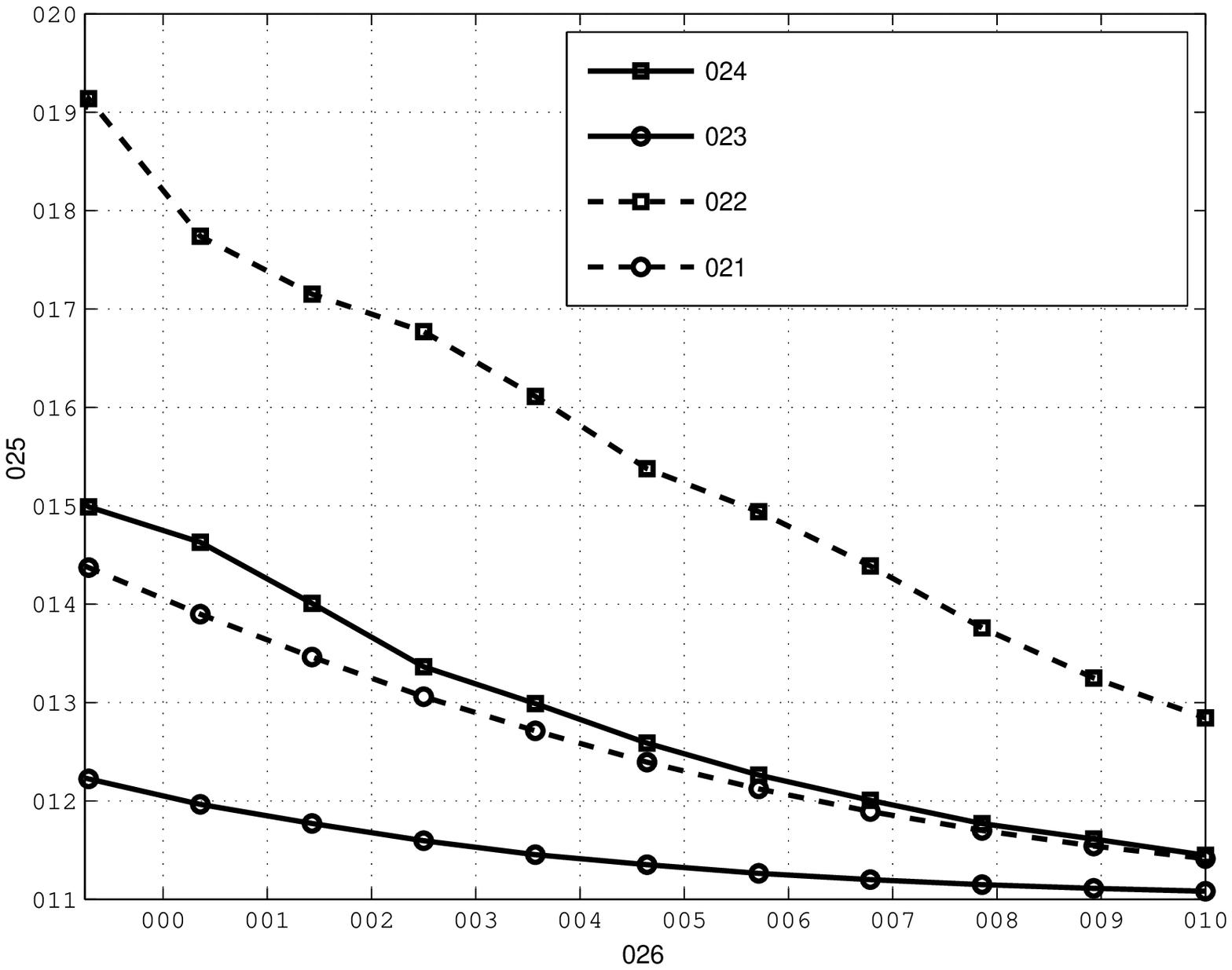}
\caption{Common rate and sum rate gap between optimal ordering and random ordering for $2$ different channel variances with $N=4$}
\label{Gvar}
\end{figure}

Fig. \ref{simFDFCR} and \ref{simFDFSR}
depict the comparison between the common rate and sum rate of the optimal ordering and random ordering
for FDF relaying in low to high SNR regimes. The upper bounds are given by max-flow min-cut
theorem \cite{cover}. Fig. \ref{GCS} illustrates the
aforementioned gap parameter and feature the effect of optimal ordering on both common rate and sum rate.
However, these figures show that the ordering effect on FDF relaying is not significant in higher SNR
regimes, as we showed earlier.  The real and imaginary parts of the channel responses during each phase are modeled by independent and identically distributed zero-mean Gaussian variables with variance $1/2$. Decreasing  this variance will increase the aforementioned gap parameters in low SNR regimes. In other words, the ordering becomes more important for higher variance of channel or in lower SNR regimes. Fig. \ref{Gvar} illustrates the gap parameter for channel realizations with variance $1$ and $\frac{1}{2}$ for $N=4$ users.

%% file: Figure2_corrected.tex
%
%
\providecommand\matlabtextA{\color[rgb]{0.000,0.000,0.000}\fontsize{6}{15}\selectfont\strut}%
\providecommand\matlabtextE{\color[rgb]{0.000,0.000,0.000}\fontsize{6}{35}\selectfont\strut}%
\providecommand\matlabtextB{\color[rgb]{0.000,0.000,0.000}\fontsize{6}{10}\selectfont\strut}%
\providecommand\matlabtextC{\color[rgb]{0.000,0.000,0.000}\fontsize{5}{10}\selectfont\strut}%
\providecommand\matlabtextD{\color[rgb]{0.000,0.000,0.000}\fontsize{5}{10}\selectfont\strut}%
\psfrag{017}[cl][cl]{\matlabtextB Random ordering, $N = 8$}%
\psfrag{018}[cl][cl]{\matlabtextB Optimal ordering, $N = 8$}%
\psfrag{019}[cl][cl]{\matlabtextB Upper bound, $N = 8$}%
\psfrag{020}[cl][cl]{\matlabtextB Random ordering, $N = 4$}%
\psfrag{021}[cl][cl]{\matlabtextB Optimal ordering, $N = 4$}%
\psfrag{022}[cl][cl]{\matlabtextB Upper bound, $N = 4$}%
\psfrag{023}[bc][bc]{\matlabtextE $S_R$ (bits of information per MWRC phase)}%
\psfrag{024}[tc][tc]{\matlabtextA ${1} / {\sigma^2}$ (dB)}%
%
%
%
\def\matlabfragNegXTick{\mathord{\makebox[0pt][r]{$-$}}}
\psfrag{000}[ct][ct]{\matlabtextC $1$}%
\psfrag{001}[ct][ct]{\matlabtextC $3$}%
\psfrag{002}[ct][ct]{\matlabtextC $5$}%
\psfrag{003}[ct][ct]{\matlabtextC $7$}%
\psfrag{004}[ct][ct]{\matlabtextC $9$}%
\psfrag{005}[ct][ct]{\matlabtextC $11$}%
\psfrag{006}[ct][ct]{\matlabtextC $13$}%
\psfrag{007}[ct][ct]{\matlabtextC $15$}%
%
%
%
\psfrag{008}[rc][rc]{\matlabtextD $0$}%
\psfrag{009}[rc][rc]{\matlabtextD $0.5$}%
\psfrag{010}[rc][rc]{\matlabtextD $1$}%
\psfrag{011}[rc][rc]{\matlabtextD $1.5$}%
\psfrag{012}[rc][rc]{\matlabtextD $2$}%
\psfrag{013}[rc][rc]{\matlabtextD $2.5$}%
\psfrag{014}[rc][rc]{\matlabtextD $3$}%
\psfrag{015}[rc][rc]{\matlabtextD $3.5$}%
\psfrag{016}[rc][rc]{\matlabtextD $4$}%
%

%% file: Figure3_corrected.tex
%
%
\providecommand\matlabtextA{\color[rgb]{0.000,0.000,0.000}\fontsize{6}{15}\selectfont\strut}%
\providecommand\matlabtextE{\color[rgb]{0.000,0.000,0.000}\fontsize{6}{35}\selectfont\strut}%
\providecommand\matlabtextB{\color[rgb]{0.000,0.000,0.000}\fontsize{6}{10}\selectfont\strut}%
\providecommand\matlabtextC{\color[rgb]{0.000,0.000,0.000}\fontsize{5}{10}\selectfont\strut}%
\providecommand\matlabtextD{\color[rgb]{0.000,0.000,0.000}\fontsize{5}{10}\selectfont\strut}%
\psfrag{016}[cl][cl]{\matlabtextB $G_S$ for $N = 8$}%
\psfrag{017}[cl][cl]{\matlabtextB $G_C$ for $N = 8$}%
\psfrag{018}[cl][cl]{\matlabtextB $G_S$ for $N = 4$}%
\psfrag{019}[cl][cl]{\matlabtextB $G_C$ for $N = 4$}%
\psfrag{020}[tc][tc]{\matlabtextA ${1} / {\sigma^2}$ (dB)}%
\psfrag{021}[bc][bc]{\matlabtextE $S_R$ (bits of information per MWRC phase)}%
%
%
%
\def\matlabfragNegXTick{\mathord{\makebox[0pt][r]{$-$}}}
\psfrag{000}[ct][ct]{\matlabtextC $1$}%
\psfrag{001}[ct][ct]{\matlabtextC $3$}%
\psfrag{002}[ct][ct]{\matlabtextC $5$}%
\psfrag{003}[ct][ct]{\matlabtextC $7$}%
\psfrag{004}[ct][ct]{\matlabtextC $9$}%
\psfrag{005}[ct][ct]{\matlabtextC $11$}%
\psfrag{006}[ct][ct]{\matlabtextC $13$}%
\psfrag{007}[ct][ct]{\matlabtextC $15$}%
%
%
%
\psfrag{008}[rc][rc]{\matlabtextD $0$}%
\psfrag{009}[rc][rc]{\matlabtextD $0.1$}%
\psfrag{010}[rc][rc]{\matlabtextD $0.2$}%
\psfrag{011}[rc][rc]{\matlabtextD $0.3$}%
\psfrag{012}[rc][rc]{\matlabtextD $0.4$}%
\psfrag{013}[rc][rc]{\matlabtextD $0.5$}%
\psfrag{014}[rc][rc]{\matlabtextD $0.6$}%
\psfrag{015}[rc][rc]{\matlabtextD $0.7$}%
%

%% file: Figure4_corrected.tex
%
%
\providecommand\matlabtextA{\color[rgb]{0.000,0.000,0.000}\fontsize{6}{15}\selectfont\strut}%
\providecommand\matlabtextE{\color[rgb]{0.000,0.000,0.000}\fontsize{6}{35}\selectfont\strut}%
\providecommand\matlabtextB{\color[rgb]{0.000,0.000,0.000}\fontsize{6}{10}\selectfont\strut}%
\providecommand\matlabtextC{\color[rgb]{0.000,0.000,0.000}\fontsize{5}{10}\selectfont\strut}%
\providecommand\matlabtextD{\color[rgb]{0.000,0.000,0.000}\fontsize{5}{10}\selectfont\strut}%
\psfrag{021}[cl][cl]{\matlabtextB FDF: $G_S$ with variance = 1/2}%
\psfrag{022}[cl][cl]{\matlabtextB FDF: $G_C$ with variance = 1/2}%
\psfrag{023}[cl][cl]{\matlabtextB FDF: $G_S$ with variance = 1}%
\psfrag{024}[cl][cl]{\matlabtextB FDF: $G_C$ with variance = 1}%
\psfrag{025}[bc][bc]{\matlabtextE $G_C$, $G_S$}%
\psfrag{026}[tc][tc]{\matlabtextA ${1} / {\sigma^2}$ (dB)}%
%
%
%
\def\matlabfragNegXTick{\mathord{\makebox[0pt][r]{$-$}}}
\psfrag{000}[ct][ct]{\matlabtextC $5$}%
\psfrag{001}[ct][ct]{\matlabtextC $6$}%
\psfrag{002}[ct][ct]{\matlabtextC $7$}%
\psfrag{003}[ct][ct]{\matlabtextC $8$}%
\psfrag{004}[ct][ct]{\matlabtextC $9$}%
\psfrag{005}[ct][ct]{\matlabtextC $10$}%
\psfrag{006}[ct][ct]{\matlabtextC $11$}%
\psfrag{007}[ct][ct]{\matlabtextC $12$}%
\psfrag{008}[ct][ct]{\matlabtextC $13$}%
\psfrag{009}[ct][ct]{\matlabtextC $14$}%
\psfrag{010}[ct][ct]{\matlabtextC $15$}%
%
%
%
\psfrag{011}[rc][rc]{\matlabtextD $0$}%
\psfrag{012}[rc][rc]{\matlabtextD $0.1$}%
\psfrag{013}[rc][rc]{\matlabtextD $0.2$}%
\psfrag{014}[rc][rc]{\matlabtextD $0.3$}%
\psfrag{015}[rc][rc]{\matlabtextD $0.4$}%
\psfrag{016}[rc][rc]{\matlabtextD $0.5$}%
\psfrag{017}[rc][rc]{\matlabtextD $0.6$}%
\psfrag{018}[rc][rc]{\matlabtextD $0.7$}%
\psfrag{019}[rc][rc]{\matlabtextD $0.8$}%
\psfrag{020}[rc][rc]{\matlabtextD $0.9$}%
%

%% file: Conclusion.tex
In this paper, we studied the effect of users' transmission ordering on the common rate and sum rate of a pairwise MWRC with FDF relaying. First, we suggested a graphical model for the data communication between the users. Then, using this model, optimal orderings were found  that maximize common rate and sum rate in the system. Moreover, we showed that for high SNR regimes, the effect of ordering becomes less important. Our claims were supported and verified by computer simulations.

%% file: ISIT.bbl
\begin{thebibliography}{10}
\providecommand{\url}[1]{#1}
\csname url@samestyle\endcsname
\providecommand{\newblock}{\relax}
\providecommand{\bibinfo}[2]{#2}
\providecommand{\BIBentrySTDinterwordspacing}{\spaceskip=0pt\relax}
\providecommand{\BIBentryALTinterwordstretchfactor}{4}
\providecommand{\BIBentryALTinterwordspacing}{\spaceskip=\fontdimen2\font plus
\BIBentryALTinterwordstretchfactor\fontdimen3\font minus
  \fontdimen4\font\relax}
\providecommand{\BIBforeignlanguage}[2]{{%
\expandafter\ifx\csname l@#1\endcsname\relax
\typeout{** WARNING: IEEEtran.bst: No hyphenation pattern has been}%
\typeout{** loaded for the language `#1'. Using the pattern for}%
\typeout{** the default language instead.}%
\else
\language=\csname l@#1\endcsname
\fi
#2}}
\providecommand{\BIBdecl}{\relax}
\BIBdecl

\bibitem{Gunduz2009}
D.~Gunduz, A.~Yener, A.~Goldsmith, and H.~Poor, ``The multi-way relay
  channel,'' in \emph{IEEE International Symposium on Information Theory
  (ISIT)}, 2009, pp. 339--343.

\bibitem{TWRC1}
B.~Rankov and A.~Wittneben, ``Spectral efficient signaling for half-duplex
  relay channels,'' in \emph{Conference Record of the Thirty-Ninth Asilomar
  Conference on Signals, Systems and Computers}, 2005, pp. 1066--1071.

\bibitem{Wilson}
M.~Wilson, K.~Narayanan, H.~Pfister, and A.~Sprintson, ``Joint physical layer
  coding and network coding for bidirectional relaying,'' \emph{IEEE
  Transactions on Information Theory}, vol.~56, no.~11, pp. 5641--5654, 2010.

\bibitem{TWRC3}
W.~Nam, S.-Y. Chung, and Y.~H. Lee, ``Capacity bounds for two-way relay
  channels,'' in \emph{IEEE International Zurich Seminar on Communications},
  2008, pp. 144--147.

\bibitem{TWRC4}
D.~Gunduz, E.~Tuncel, and J.~Nayak, ``Rate regions for the separated two-way
  relay channel,'' in \emph{46th Annual Allerton Conference on Communication,
  Control, and Computing}, 2008, pp. 1333--1340.

\bibitem{Popovski}
P.~Popovski and H.~Yomo, ``The anti-packets can increase the achievable
  throughput of a wireless multi-hop network,'' in \emph{IEEE International
  Conference on Communications (ICC)}, vol.~9, 2006, pp. 3885--3890.

\bibitem{Moslem}
M.~Noori and M.~Ardakani, ``Optimal user pairing for asymmetric multi-way relay
  channels with pairwise relaying,'' \emph{IEEE Communications Letters},
  vol.~16, no.~11, pp. 1852--1855, 2012.

\bibitem{Ong_Binary}
L.~Ong, S.~Johnson, and C.~Kellett, ``An optimal coding strategy for the binary
  multi-way relay channel,'' \emph{IEEE Communications Letters}, vol.~14,
  no.~4, pp. 330--332, 2010.

\bibitem{Cadambe}
V.~Cadambe, ``Multi-way relay based deterministic broadcast with side
  information: Pair-wise network coding is sum-capacity optimal,'' in
  \emph{46th Annual Conference on Information Sciences and Systems (CISS)},
  2012, pp. 1--3.

\bibitem{Ong}
L.~Ong, S.~Johnson, and C.~Kellett, ``The capacity region of multiway relay
  channels over finite fields with full data exchange,'' \emph{IEEE
  Transactions on Information Theory}, vol.~57, no.~5, pp. 3016--3031, 2011.

\bibitem{Katti}
S.~Katti, H.~Rahul, W.~Hu, D.~Katabi, M.~Medard, and J.~Crowcroft, ``{XOR}s in
  the air: Practical wireless network coding,'' \emph{IEEE/ACM Transactions on
  Networking}, vol.~16, no.~3, pp. 497--510, 2008.

\bibitem{Nam2010}
W.~Nam, S.-Y. Chung, and Y.~H. Lee, ``Capacity of the {G}aussian two-way relay
  channel to within $\frac{1}{2}$ bit,'' \emph{IEEE Transactions on Information
  Theory}, vol.~56, no.~11, pp. 5488--5494, 2010.

\bibitem{cayley}
A.~Cayley, ``A theorem on trees,'' \emph{Quart. J. Math}, vol.~23, no. 376-378,
  p.~69, 1889.

\bibitem{cover}
T.~M. Cover and J.~A. Thomas, \emph{Elements of Information Theory}.\hskip 1em
  plus 0.5em minus 0.4em\relax New York, NY, USA: Wiley-Interscience, 1991.

\end{thebibliography}
